# How to Align Large Language Models for Teaching English? Designing and Developing LLM based-Chatbot for Teaching English Conversation in EFL, Findings and Limitations


Jaekwon Park[1], Jiyoung Bae[1], Unggi Lee[1, 4]†, Taekyung Ahn[1], Sookbun Lee[1], Dohee Kim[1], Aram Choi[1]

Yeil Jeong[2], Jewoong Moon[3]†, Hyeoncheol Kim[4]†

Enuma, Inc.[1], Seoul National University[2], The University of Alabama[3], Korea University[4]

{Jaekwon, jiyoung}@enuma.com, codingchild@korea.ac.kr,
{taekyung, blackdew, dohee, ahram}@enuma.com,
yell001@snu.ac.kr, jmoon19@ua.edu, harrykim@korea.ac.kr



## Abstract

This study investigates the design, development, and evaluation of a Large Language Model (LLM)-based chatbot for teaching English conversations in an English as a Foreign Language (EFL) context. Employing the Design and Development Research (DDR), we analyzed needs, established design principles, and iteratively refined a chatbot through experimenting various LLMs and alignment methods. Through both quantitative and qualitative evaluations, we identified the most effective LLM and its prompt combination to generate high-quality, contextually appropriate responses. Interviews with teachers provided insights into desirable system features, potential educational applications, and ethical considerations in the development and deployment of the chatbots. The design iterations yielded the importance of feedback mechanisms and customizable AI personas. Future research should explore adaptive feedback strategies, collaborative approaches with various stakeholders, and the integration of insights from human-computer interaction (HCI) and user experience (UX) design. This study contributes to the growing body of research on applying LLMs in language education, providing insights and recommendations for the design, development, and evaluation of LLM-based chatbots for EFL conversation practice. As the field evolves, ongoing research and collaboration among educators, AI engineers, and other stakeholders will be essential to harness the potential of these technologies to enhance language learning experiences.

*Keywords: Large language models, Alignment, English as a Foreign Language, Chatbot, Design and Development Research*




## Introduction

In English as a Foreign Language (EFL) education, conventional methodologies for teaching conversation skills require extensive resources and significant time investment from educators and learners (Bagheri & Mohamadi Zenouzagh, 2021; Yaman, 2014). Despite the availability of diverse resources, these approaches often fail to effectively simulate real-life conversational contexts, limiting learners' opportunities for spontaneous and natural dialogue practice (Asaad et al., 2020). Without ample opportunities to practice conversations in authentic situations, EFL learners could become adept at the formal aspects of English yet appear ill-prepared for everyday communication (Anagnostopoulou et al., 2023).

Chatbot technology has introduced an innovative alternative for facilitating English conversation practice. Initially, rule-based chatbots were employed; however, their limited capability to generate human-like responses significantly hampered their effectiveness as educational tools (Yang et al., 2022). This limitation led to further explorations of more advanced solutions, namely the implementation of Large Language Models (LLMs), such as ChatGPT, which have demonstrated proficiency in generating human-like language responses (Adeshola & Adepoju, 2023; Lee et al., 2024; OpenAI, 2021).

LLMs offer a promising avenue for enhancing EFL conversation instruction due to their capability to provide immediate, contextually relevant, and varied responses customized to each learner's inputs (Al-Graady & Mahyoob, 2023; Han et al., 2023; Lee et al., 2023). This capability to engage students in spontaneous yet coherent dialogues aligns well with communicative language teaching principles that focus on contextualized and meaningful interactions (Kim et al., 2023; Young & Shishido, 2023). By simulating real-life conversational scenarios, LLM-based chatbots can create an engaging and personalized learning environment for



developing conversational fluency, a critical challenge that has proven difficult to address (Kasneci et al., 2023; Kirk et al., 2023).

However, the development of LLM-based chatbots for EFL faces challenges in aligning the models with specific educational requirements, both ensuring the generation of pedagogically valid content and evaluation of their effectiveness in promoting language learning (Al-Garaady & Mahyoob, 2023; Meyer et al., 2023). While recent studies have explored the potential of LLMs in language education (Jeon & Lee, 2023; Kasneci et al., 2023), there remains a paucity of design and development research explicitly illustrating the design journey of LLM-based chatbots in authentic EFL settings.

Moreover, the integration of LLMs in educational contexts is a significant consideration. Data privacy is a concern when students' inputs are processed externally, and the substantial computational resources required may render many LLM solutions prohibitively expensive for schools and institutions (Huallpa, 2023; Rawas, 2023; Tu et al., 2023; Yan et al., 2024). An alternative approach accordingly involves customizing smaller language models explicitly tailored to EFL curricula, which can be locally deployed to mitigate privacy risks (Alawida et al., 2023; Whalen & Mouza, 2023; Wu et al., 2023). The present study aims to investigate and improve the optimization and evaluation of LLMs for application to EFL conversation teaching by exploring the following key questions:

- RQ1: Which alignment method optimizes LLMs for EFL conversation teaching?

- RQ2: How can LLMs be effectively evaluated in teaching English conversation within EFL settings?

- RQ3: What are teachers' perceptions and responses to implementing LLMs in EFL conversation instruction?



To address these questions, we employed the Design and Development Research (DDR) methodology (Richey & Klein, 2005; Richey & Klein, 2014), which provides a structured framework for designing, evaluating, and refining educational technology and software. Under DDR, we conducted a needs analysis, established design principles, and fine-tuned chatbot prototypes.

## Literature Review

### Large Language Model-based chatbot in English Education

Integrating LLMs into EFL education through chatbots has been a novel approach to foster students' linguistic interactions, especially in conversational contexts. Recent advancements leverage the progress in conversational AI (Hsu et al., 2023) to offer learners highly-engaging experiences akin to real-life conversations, which are crucial for communicative language teaching (CLT) that prioritizes practical language use in authentic scenarios (Sha, 2009). The foundational efforts in creating virtual human dialogues, such as the NPCEditor (Leuski & Traum, 2011) and the Virtual Human Toolkit (Hartholt et al., 2013), have paved the way for developing conversational chatbots.

The pedagogical foundation for employing LLM-based chatbots in EFL draws from insights across several pivotal studies. The embodied conversational agents (Cassell, 2000) highlight the importance of simulating face-to-face interaction, while the collaborative storytelling capabilities of tools like "Storybuddy" (Zhang et al., 2022) illustrate the chatbots' potential to facilitate interactive and engaging learning scenarios. The design and deployment of these chatbots also incorporate the principles of just-in-time feedback and scaffolding, mimicking effective human tutors within a CLT framework, as demonstrated in the adaptive



conversational aids aligned with the core design principles of intelligent tutoring systems (Graesser et al., 2001; VanLehn, 2006).

The advent of LLMs like ChatGPT and their capacity to generate human-like text and engage in complex dialogues (Bang et al., 2023) have broadened the scope of interactive language learning activities, supporting the CLT focus on meaningful communication (Yang et al., 2023; Sallam, 2023). However, despite the growing interest in LLM-based chatbots for language learning, there still needs to be more research specifically investigating their design, development, and evaluation in authentic EFL settings (Eisenring et al., 2024; Riza et al., 2023).

The integration of LLM-based chatbots into EFL education, informed by previous research, including studies on conversational dynamics (Cole et al., 1999; Jurafsky et al., 2009) and multimodal dialogue systems (Allen et al., 2019), presents a promising path for enhancing language learning through CLT. This integrated approach aims to deliver real-life communication and learner motivation while emphasizing the role of feedback in the language learning process.

## Alignment Method for Large Language Model

LLMs have profoundly impacted text generation, predicting and generating the next word. They have been trained using massive amounts of data to secure good LLM performance. However, the bias, toxicity, and privacy issues inherent in training data have yet to be fully addressed, which means that LLMs can generate unintended content. Consequently, an alignment method is essential to generate the desired answers without either harming or confusing the user (Zhao et al., 2023) by tuning coherent topics. These alignment issues have allowed researchers to explore various techniques to improve the controllability and predictability of LLM behaviors (Ouyang et al., 2022; Wei et al., 2021).



**Instruction Tuning**

One of the ways to achieve alignment is through Instruction Tuning, which consists of a dataset of the form (INSTRUCTION, OUTPUT), consisting of human instructions to the model and the desired output according to the INSTRUCTIONS (Zhang et al., 2023). This data format enables more controllable and predictable model behaviors compared to standard LLMs. It can be adapted to specific domains without requiring extensive retraining or architectural changes. Instruction Tuning has improved LLM performance on various tasks, including question-answering, summarization, and dialogue generation (Shu et al., 2024; Zhang et al., 2023).

**Prompting Strategies**

In addition to Instruction Tuning, researchers have explored various prompting strategies to align LLM behavior with specific tasks and requirements. Two prominent approaches are in-context learning (ICL) and chain-of-thought prompting (CoT). First, ICL relies on providing h task-specific examples in the prompt to guide the LLM's output without explicit fine-tuning (Brown et al., 2020). The prompt begins with a task description, selects examples from the task dataset as demonstrations, and adds test instances to these demonstrations. These are then used as input for the LLM to generate output. LLM can recognize and perform new tasks without explicit gradient updates based on the task demonstration.

CoT reinforces in-context learning by incorporating intermediate reasoning steps into the prompt (Chu et al., 2023; Wei et al., 2022). This prompting strategy aims to increase LLM performance on complex reasoning tasks, such as arithmetic reasoning (Miao et al., 2020), commonsense reasoning (Talmor et al., 2019), and symbolic reasoning (Wei et al., 2022). While ICL organizes prompts into input-output pairs, CoT integrates intermediate reasoning steps into



the prompts that lead to the final output. Recent studies have shown that CoT prompting significantly improves  LLM performance across a variety of reasoning tasks compared to ICL (Kojima et al., 2022; Zelikman et al., 2022).

**Evaluation Method for Large Language Model**

The complexity of LLMs, coupled with the significant costs associated with training them—costs that scale with model size—makes it challenging to precisely refine these models in a desired direction and impractical to conduct repetitive training sessions until the desired performance is achieved. As a result, ongoing discussions have focused on evaluating LLMs' performance (Guo et al., 2023).

Evaluation of LLMs typically involves assessing them based on various aspects, such as knowledge, reasoning, tool learning, toxicity, truthfulness, robustness, and privacy. As LLMs become proficient in acquiring and applying extensive knowledge, and the boundaries of their capabilities in downstream tasks become ambiguous, evaluations are increasingly conducted based on various comprehensive capabilities (Guo et al., 2023).

There are both qualitative and quantitative methods for evaluating LLMs. Qualitative evaluation involves human evaluators, while quantitative assessments utilizes credible benchmarks comprising questions selected from verified or well-known tests, such as MMLU (Hendricks et al., 2021b), HELM (Liang et al., 2022), BIG-bench (Srivastava et al., 2022), and Chatbot Arena (Zheng et al., 2023). Researchers have also introduced automated evaluation systems that mimic the qualitative assessment approach of human evaluators using LLMs, such as GPT-4, which have demonstrated matching or surpassing human performance in several tasks (Chang et al., 2023).



In addition, domain-specific evaluation methods exist, such as PubMedQA (Jin et al., 2019) and MultiMedQA (Singhal et al., 2022) in the medical field, as well as benchmarks tailored for the legal, programming, and financial domains (Guo et al., 2023).

Despite the potential applications of LLMs in education, there is a lack of widely accepted benchmarks or evaluation methods (Chang et al., 2023). Recent studies have attempted to close this gap by exploring the use of LLMs in generating educational content, providing personalized feedback to learners (Chang et al., 2023), and automated essay scoring and feedback generation (Mizumoto & Eguchi, 2023).

The present study contributes to this emerging research area by developing and demonstrating a comprehensive evaluation framework for assessing the performance and effectiveness of LLMs in the context of an English conversation chatbot for EFL learners. Through a combination of quantitative and qualitative evaluation methods, the study aims to advance LLM evaluation in the educational domain and provide insights to inform future research and development efforts. It necessitates empirical design and development studies regarding LLM-based chatbot evaluation in education.

## Method

In this study, we employed the Design and Development Research (DDR) methodology (type 1) (Richey & Klein, 2005; 2014). DDR is particularly effective for developing and validating educational technologies. This method involves a systematic investigation of design, deployment, and evaluation processes to produce both instructional and non-instructional evidence grounded in empirical evidence. The cyclic nature of DDR makes it an ideal framework for the evidence-based development of a chatbot leveraged by LLMs.



## Research Context

The development of an LLM-based chatbot in this study was part of the larger AI Digital Textbook (AIDT) project, launched by South Korea's Ministry of Education to promote personalized learning via AI technology. The AIDT project aims to develop interactive digital learning resources that adapt to learners' needs and are aligned with the curriculum. Although AI chatbots are not explicitly required by the AIDT project, the national English curriculum underscores the use of interactive digital tools, such as chatbots, to enhance English communication skills—a core goal of the curriculum (MOE, 2022). By providing task-based language practice that simulates real-life interactions, the development of this LLM-based chatbot directly supports the curriculum's focus on developing communicative competence through engaging, contextual-rich experiences.

## Procedure

The research procedure unfolded in six phases. Phase 1 involved conducting a needs analysis through interviews with product owners (POs) to identify objectives, specific requirements, and constraints. In Phase 2, these findings were synthesized to establish design principles for developing the LLMs. Phase 3 investigated state-of-the-art LLMs and training methods, utilizing a literature review and market analysis. In Phase 4, a dataset was designed to fine-tune the selected LLMs, evaluation criteria and prompt templates were developed using both base LLMs (n = 5) and fine-tuned LLMs (n = 5).  Phase 5 involved a comprehensive evaluation of these LLMs, employing  a combination of 17 LLMs, including GPT-4, to objectively assess their performance using the established templates (n = 5). As the final phase, Phase 6, involved conducting interviews with three teachers to gather qualitative feedback on the usability of the third prototype. This data was analyzed using a qualitative coding method (Pike, 2015) to



identify key themes and insights that elucidate the chatbot's usability and potential to support

English language learning.

**Figure 1**
*Research Procedure*

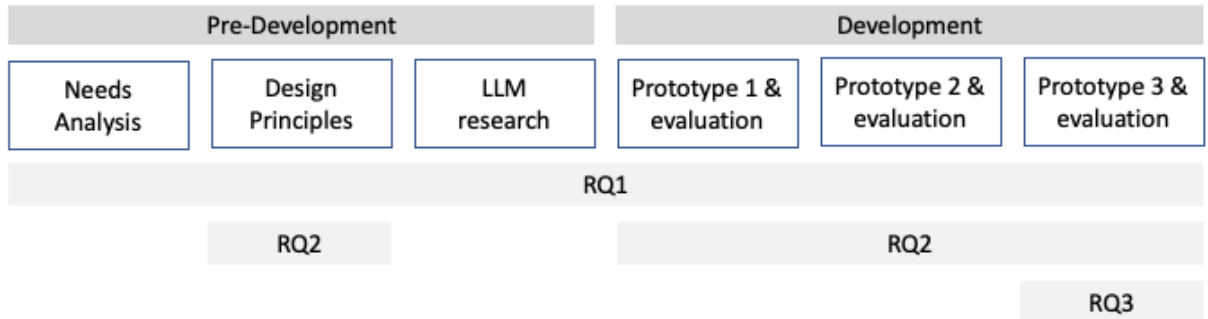

**Participants, Data Collection and Analysis**

We interviewed subject matter experts (SMEs) in elementary education to investigate the

usability, strengths, weaknesses, and perspectives on teacher analytics. The SMEs were selected

based on their proficiency in English education and professionalism with AI technologies (Table

1). We conducted open coding of the expert interviews to discover insights into the usability and

effectiveness of the developed English chatbot. Prior to the interviews, we obtained informed

consent from the participating teachers and provided them with a detailed explanation of the

research procedures, ensuring compliance with ethical research practices and the study protocol

approved by the Institutional Review Board (IRB).

**Table 1**
*The Information of the Interviewees*

| Interviewee | Gender | Teaching Experiences | Grade Level | Major |
|:---:|:---:|:---:|:---:|:---:|
| A1 | Male | 8 years | Elementary school | Elementary education |



| A2 | Female | 9 years | High School | English education |
|----|--------|---------|-------------|-------------------|
| A3 | Male | 8 years | Elementary school | Elementary education |

After analyzing expert interviews, we identified seven critical themes subdivided into twelve categories, as illustrated in Table 2. This code table organizes the insights from the interviews into specific areas such as feature modifications, educational use, ethical considerations, persona development, AI-human interaction, and engagement strategies. These categories facilitate a focused interpretation of the data, extracting actionable insights for improvement. Areas for Improvement include enhancements to chatbot functionalities and features for increased utility. Educational Use involves utilizing the chatbot to support learning through conversational practice. Functionality ensures the chatbot corrects errors and responds in a timely, contextually relevant manner. Ethical Considerations focus on adhering to ethical standards and avoiding harmful content. Persona involves customizing the chatbot's personality to improve relatability and interaction quality. AI-Human Interaction aims to develop a user-friendly interface that fosters natural interactions. Feedback emphasizes integrating user feedback mechanisms to personalize learning experiences.

**Table 2**
*Code Table*

| Theme | Category |
|-------|----------|
| Areas for Improvement | Feature Modification |
|  | Additional Feature |
| Educational Use | Usage Situations |
| Functionality | Pointing out Inappropriate Responses |
|  | Immediate and Contextual Responses |
| Ethical Considerations | Removal of Harmful or Inappropriate Content |
| Persona | Customizable Persona |
|  | Stereotyped and Unnatural Responses |



| AI-Human Interaction | Familiarity |
|---|---|
| | Natural Feedback during Conversations |
| Feedback | Learner-Selected Feedback |
| | Maintaining Engagement in Conversation |

**Experiment Setting**

*Large Language Models*

This research employed a variety of LLMs to develop and assess a chatbot designed for English conversation practice tailored to EFL students. We compared models including GPT-4, GPT-3.5, and two variants of Llama-2 (13b and 70b), selected based on factors like size, training data, and architecture. The study also assessed the feasibility of using open-source LLMs, such as Llama-2, as cost-effective alternatives to proprietary models like GPT-4 and GPT-3.5, evaluating their effectiveness in the given educational context.

*Alignment Methods*

We explored three primary alignment methods to tailor the LLMs to the needs of an English conversation chatbot for EFL learners: fine-tuning, instruction-tuning, and prompting. Fine-tuning adjusts pre-trained LLMs using a dataset of English conversations that reflects the chatbot's intended interactions, encompassing a diverse range of conversation topics, learner proficiency levels, and potential challenges. Instruction-tuning focuses on optimizing the LLMs' ability to follow instructions by training on a dataset that pairs explicit instructions with corresponding outputs. Prompting, or in-context learning, involves giving the LLM with examples of desired behaviors within the input prompt, enabling the model to adapt its responses to the context. This approach tested various prompt structures to identify the most effective. These methods were assessed to optimize the chatbot's conversational abilities, enhance its



language learning utility for EFL learners, and inform the iterative development and refinement of the chatbot according to established design principles.

### *Evaluation Methods*

We employed two primary evaluation methods, quantitative and qualitative, to assess the LLMs in the development of an English conversation chatbot for EFL learners. The quantitative evaluation leveraged the advanced capabilities of GPT-4 to measure the performance of other LLMs, employing a set of metrics and criteria established in Phase 2 (Table 3). These metrics assessed the quality, coherence, and pedagogical value of the conversations generated by each LLM, enabling a standardized comparison of their performance. Conversely, the qualitative evaluation involved gathering insights and feedback from domain experts—precisely, three English teachers—to gauge the chatbot's practical usability and educational value. Through semi-structured interviews, where teachers interacted with the chatbot's final l prototype, their insights were gathered and analyzed using qualitative coding methods to identify common themes that inform the study's discussion and implications.

**Table 3**
*Quantitative Evaluation Metrics and Criteria*

| Metric | Criteria | Related principles |
|---|---|---|
| Contextual coherence | Maintains context and relevance throughout the conversation | 2a, 2c, 2e |
| Grammatical accuracy | Generates grammatically correct responses | 3b |
| Conversational feedback quality | Provides appropriate conversational feedback to guide the user and maintain the flow of the conversation | 1a, 2b |
| Instructional | Offers accurate and constructive feedback on language use and | 2a, 3b |



| feedback quality | errors | |
| Conversation length adherence | Adheres to the specified conversation length and topic | 2e |

## Results

**Figure 2**

*Phase of Design Journey and Design Principles*

| Phase | Phase 1 | Phase 2 | Phase 3 | Phase 4 | Phase 5 | Phase 6 |
|---|---|---|---|---|---|---|
| **Phase Name** | Need Analysis | Design principles | Exploratory analysis about LLMs | First prototype and evaluation | Second prototype and evaluation | Design principles |
| **Research Question** | RQ1, 3 | RQ1, 3 | RQ1, 3 | RQ1, 2 | RQ1, 2 | RQ3 |
| **Details** | Needs Analysis | Made design principles (direction of alignment and evaluation) | Performance Test and Alignment | Alignment, Evaluation | Alignment, Evaluation | Evaluation, Interview analysis of 3 teachers |
| **User Experience** | 1a | | | | | |
| **Functionality** | 2a | | | | | |
| | 2b | | | | | |
| | 2c | | | | | |
| | 2d | | | | | |
| | 2e | | | | | |
| **Compliance with Guidelines** | 3a | | | | | |
| | 3b | | | | | |

**Design Principle**

## Phase 1: Needs Analysis

In the first phase, we conducted a needs analysis with product owners (POs) who are interested in developing an LLM-based chatbot to facilitate English conversation practice among elementary EFL students. Interviews with POs led to the identification and categorization of design guidelines for the chatbot into three main areas: (1) User experience requirements, focusing on creating a smooth and engaging conversational experience; (2) Service functionality requirements, addressing the core functionality and behaviors of the chatbot; and (3) Compliance



requirements, ensuring the chatbot adheres to digital textbook standards and is appropriate for classroom use.

**Phase 2: Design Principles**

In Phase 2, following the needs analysis, we established design principles for the development of the LLM-based English conversation chatbot. These design principles were categorized into three main areas: user experience, functionality, and compliance with guidelines. User experience principles focused on creating a natural, engaging, and personalized conversation experience for the learners. Functionality principles ensured the chatbot provided effective task-based language teaching, clear, conversational turn-taking, and appropriate topic management. Lastly, compliance principles addressed the need to adhere to the AI digital textbook guidelines and maintain content appropriateness for classroom settings. These principles are detailed in Table 4.

**Table 4**
*Design Principles*

| Category | Design Principle |
|---|---|
| User Experience | 1a. *Diverse and Engaging Responses*: The LLM should generate varied and natural responses to avoid a machine-like conversation experience. |
| Functionality | 2a. *Task-Based Language Teaching*: The chatbot should facilitate free conversation on topics selected from English textbooks for Korean elementary students, aligning with the principles of task-based language teaching. |
| | 2b. *Clear Conversational Turn-Taking*: The chatbot should initiate conversations with questions and generate dialogue that clearly signals the beginning, end, and turns of the conversation, particularly benefiting novice learners. |
| | 2c. *Topic Guardrail*: When students input off-topic text, the chatbot should acknowledge the input and gently steer the conversation back to the main topic.<br>- Example |
|  | - Student: I like baseball<br>- Chatbot: Oh, baseball is fun, but we were talking about weather. Do you like sunny days? |
|  | 2d. *User Intent Recognition*: The LLM should recognize users' intent to end or leave the conversation, responding appropriately without prolonging the interaction. |
|  | 2e. *Appropriate Conversation Length*: The conversation should conclude within 5 to 10 turns of dialogue, based on typical textbook conversations, while allowing for a buffer of 2 to 3 turns for off-topic talk and redirection. |
| Compliance with Guidelines | 3a. *LLM Selection*: Due to constraints of AI digital textbook guidelines, the chatbot should be developed using open-source LLMs with commercially allowable licenses, rather than closed LLMs like GPT-4 or BARD. |
|  | 3b. *Content Appropriateness*: The dialogue generated by the chatbot should align with the conversational level typically found in the textbooks of the target users, specifically 3rd and 4th grade Korean elementary students, to ensure its suitability for use in AI digital textbooks. |

## Phase 3: Investigating the State-of-the-art  LLMs and Training Method

Assessing the performance and capabilities of existing LLMs was essential, given the need for

reliable data on optimal techniques in this rapidly advancing field. We conducted experiments to

evaluate the conversational abilities and adaptability of GPT-4 and LLama2 70b using few-shot

prompts, aligning with design principles 1a, 2a, 2e. While both models demonstrated basic

conversational capabilities, LLama2 70b often generated excessively long sentences and

complex vocabulary. In a test requiring communication akin to conversing with a 5-year-old,

GPT-4 adapted its language appropriately, while LLama2 70b's responses remained complex and

verbose. These observations informed further evaluations and development strategies.

Subsequent experiments compared the original Llama-2 70b with itsGPTQ variant; the

original model excelled in handling JSON formatting, while the GPTQ version produced more

natural conversations, although the underlying reasons for this difference were unclear. To

enhance training processes, we employed GPT-4 to generate a large dataset of diverse dialogue



samples, resulting in approximately 21,000 samples across seven objectives. Fine-tuning with this dataset indicated potential improvements but also surfaced some issues like repetitive questioning by Llama-2.

Concluding this phase, we identified the necessity to establish comprehensive evaluation criteria for conversational AI. Clear metrics and benchmarks were outlined for language understanding, response relevance, coherence, and task completion, enabling systematic performance evaluation and targeted model enhancements.  Moving forward, we planned to develop a robust evaluation framework and iteratively fine-tune the models, leveraging the insights gained from these experiments to enhance the naturalness, context-awareness, and task orientation of conversational AI assistants.

**Phase 4: First Prototype and Evaluation (RQ1 & RQ2)**
***Generate Synthetic Dataset and Training Language Models***
In the fourth phase, we developed the first prototype of the LLM-based English conversation chatbot, utilizing open-source LLMs to adhere to cost and licensing requirements (Design principle 3a). The first prototype focused on the "Functionality principle," from Phase 2, integrating essential features while maintaining natural conversation flow.

The development involved creating a fine-tuning dataset tailored to POs' needs identified in Phase 1. Three key factors influenced the dataset design. First, we generated virtual dialogue data representing speakers with varying English proficiency levels, drawing on previous research in virtual dialogue generation (Bae et al., 2022; Li et al., 2023). This ensures the language model can effectively interact with users of different language skills. Second, to promote diversity and prevent biases, we varied the data within each category, as using too similar datasets can cause models' biases that favor certain responses (Mukherjee et al., 2023). Third, we addressed the



issue of repetitive responses in extended conversations by incorporating  previous dialogue

contexts in data generation, allowing the model to engage in varied conversations.

Leveraging GPT-4, we created a synthetic dataset of 3,000 dialogue samples with seven

tasks based on these considerations. This dataset, versioned as v1, forms the foundation of the

prototype and is designed to meet the Phase 1 requirements of POs by fine-tuning the language

models to better understand and engage with users across a wide range of English proficiency

levels and dialogue contexts. To evaluate the effect of model size on performance, we selected

five base LLMs (LLama2-7b, 13b, 70b, Mistral-7b, and zephyr-7b) and fine-tuned them using

the 3,000 synthetic datasets. The phase concluded with five base and five fine-tuned LLMs ready

for subsequent evaluation.

### *Evaluation*

Our qualitative evaluation framework employed six criteria, each aligned with specific design

principles, to assess the performance of LLMs. The framework utilized targeted prompts to test

LLMs' capabilities across various conversational scenarios, detailed in Table 4 and Appendix 2.

The first criterion evaluated the LLMs' ability to initiate conversations using 'First Questions'

and 'Unlearned Topic Questions' prompts, examining adherence to conversational directives

(Design principle 2a). . The second criterion assessed the LLMs' responses to exit commands

through 'User Early Termination', 'User Termination', and 'System Termination' prompts,

focusing on their ability to terminate the conversation as requested (Design principle 2d). The

third criterion examined how well the LLMs handled irrelevant or off-topic content using

'Irrelevant Answer' prompts (Design principle 2c). The fourth criterion assessed conversation

flow management with 'Preventing Question' prompts (Design principle 2e). The fifth criterion



looked at the LLMs' ability to recognize and respond to user names, and the sixth focused on their handling of inappropriate or toxic content through 'Response to Profanity' prompts. This comprehensive approach provided insights into the LLMs' performance in initiating , adhering to commens, handling irrelevant and inappropriate content, managing conversation flow, and personalizing responses.

**Table 4**

*Pairs of criteria and prompt type for assessment (details in the Appendix 2)*

| Criteria | Prompt Type for Assessment |
|---|---|
| Ability to initiate conversation based on a given prompt | • First Questions<br>• Unlearned Topic Questions |
| Understanding of exit commands and their ability to shutdown upon a specific request | • User Early Termination<br>• User Termination<br>• System Termination |
| Ability to handle irrelevant or off-topic content | • Irrelevant Answer |
| Ability to manage the flow of the conversation | • Preventing Question |
| Basic understanding and response appropriateness | • Name Recognition |
| Handling of inappropriate or toxic content | • Response to Toxicity |

Using a designated assessment prompt, we evaluated a total of ten LLMs, including GPT-4 (Chang et al., 2023), with each model being tested three times. The results, presented in Tables 5 and 6, detail the quantified qualitative performance of the LLMs, with a comprehensive breakdown available in Appendix 3. Our evaluation revealed that Mistral-7B, as shown in Table 5, emerged as the most suitable model for our assessment criteria. This finding suggests a



strategic shift towards focusing on prompt engineering, optimizing our resources by reducing the need for extensive model fine-tuning.

**Table 5**
*Quantified qualitative performance of LLMs with no additional training (details in the Appendix 3)*

| Criteria | LLMs with no additional training | | | | |
|---|---|---|---|---|---|
| | Llama-7b | Llama-13b | Llama-70b | Mistral-7b | Zephyr-7b |
| First Questions | 5.0 | 5.0 | 5.0 | 5.0 | 5.0 |
| Unlearned Topic Question | 5.0 | 1.0 | 2.0 | 5.0 | 5.0 |
| User Early Termination | 1.0 | 1.0 | 1.0 | 5.0 | 4.5 |
| User Termination | 1.0 | 5.0 | 4.0 | 3.0 | 5.0 |
| System Termination | 1.0 | 1.0 | 1.0 | 5.0 | 1.0 |
| Irrelevant Answer | 1.5 | 5.0 | 4.333 | 5.0 | 4.5 |
| Preventing Question | 4.0 | 4.0 | 2.6 | 4.8 | 4.0 |
| Name Recognition | 3.4 | 2.886 | 3.7 | 3.2 | 2.9 |
| Response to Toxicity | 5.0 | 5.0 | 5.0 | 5.0 | 5.0 |
| Average Score | 3.16 | 3.63 | 3.37 | **4.37** | 3.95 |

**Table 6**
*Quantified qualitative performance of fine-tuned LLMs (details in the Appendix 3)*

| Criteria | Fine-tuned LLMs | | | | |
|---|---|---|---|---|---|
| | Llama-7b | Llama-13b | Llama-70b | Mistral-7b | Zephyr-7b |
| First Questions | 5.0 | 5.0 | 5.0 | 5.0 | 5.0 |
| Unlearned Topic Question | 5.0 | 1.0 | 2.33 | 5.0 | 5.0 |



**Table 6**
*Quantified qualitative performance of fine-tuned LLMs (details in the Appendix 3)*

| | | | | | |
|---|---|---|---|---|---|
| User Early Termination | 1.0 | 1.0 | 1.0 | 5.0 | 4.17 |
| User Termination | 2.33 | 3.93 | 3.35 | 3.67 | 4.9 |
| System Termination | 1.0 | 1.0 | 1.0 | 1.0 | 5.0 |
| Irrelevant Answer | 3.843 | 4.827 | 4.873 | 4.927 | 4.5 |
| Preventing Question | 3.7 | 4.616 | 4.238 | 3.926 | 3.925 |
| Name Recognition | 4.064 | 3.748 | 3.982 | 3.466 | 3.708 |
| Response to Toxicity | 4.83 | 5.0 | 4.47 | 4.73 | 4.78 |
| Average Score | 3.58 | 3.85 | 3.82 | 3.85 | 4.01 |

## Phase 5: Second Prototype and Evaluation (RQ1 & RQ2)

### Prompt Engineering

In Phase 5, we concentrated on optimizing LLMs to our design principles through prompt engineering. After iterative testing to find effective prompts, we developed five versions of prompt templates, each aimed at fulfilling the functionality principles. Table 7 includes a summary of the first prompt template with the complete templates detailed in Appendix 4.

**Table 7**
*Example of Prompt Template Version 1 (details in Appendix 4)*

| Version | Short Version of Prompt Templates |
|---|---|
| v1 | Role: You are an EFL conversation practice bot. Your job is to have a chat with the user to help the user achieve the objective and use the key expressions. Our students are 10 year old children. You must be kind to children. Use simple words for kids.<br><br>You must start the conversation with a question and wait for the user input. Your response to the user must refer to the below examples. |



**Table 7**

*Example of Prompt Template Version 1 (details in Appendix 4)*

---

…

Create a hint for user input with your message, make JSON output as following:
{
…
}

GOOD_EXAMPLE_SENTENCEs : ..
GOOD_WORDs : ..
You cannot close before the user says that he/she wants to close the conversation.
When the user says that he/she wants to close the conversation, you must set true to `is_finished` field on the last message.

---

### Evaluation

For evaluation, we selected 17 LLMs from Hugging Face and paired them with the five prompt templates, creating 85 unique LLM-prompt combinations. , We assessed these combinations 19 times each using the criterion prompt from Phase 4. The evaluation process involved a quantitative analysis where each combination was rated on (a) Relevance, (b) Coherence, and (c) Adherence to conversational guidelines. A standardized rubric was used to assign numerical ratings, ensuring objectivity and consistency. These scores were then aggregated to determine an overall performance score for each LLM-prompt combination.

Additional metrics such as error rate and estimated operation times were also considered to identify the most efficient and reliable models. The error rate measured the frequency of errors or inconsistencies in the generated responses, and the estimated time assessed the computational efficiency of each combination. This comprehensive approach helped us to identify the best solution for real-world educational applications.



The results, compiled in Table 8 and Appendix 5, highlighted the "dolphin-2.6-mixtral-8x7b" model paired with the "prompt-v5" template (score 4.41) as the highest scorer, indicating its ability to generate high-quality and contextually-appropriate responses. However, the "NeuralHermes-2.5-Mistral-7B-AWQ" model with the "prompt-v1" template demonstrated the best balance among performance (score of 4.35), lower error rate (0.000) and reduced estimated time (5.34). Based on these findings, we decided to advance tthis model and prompt combination to the third prototype, aiming for a balance of performance, reliability, and efficiency optimal for real-world deployment.

**Table 8**
*Top5 quantified qualitative evaluation performance of LLMs. We chose NeuralHermes-2.5-Mistral-7B-AWQ with prompt-v1, considering the balance of score, error and time.*

| LLMs | Prompt version | Score | Error | Etime |
|---|---|---|---|---|
| dolphin-2.6-mixtral-8x7b | V5 | **4.41** | <u>0.0526</u> | 38.40 |
| **NeuralHermes-2.5-Mistral-7B-AWQ** | V1 | <u>4.35</u> | **0.000** | 5.34 |
| OpenHermes-2.5-Mistral-7B-AWQ | V2 | 4.31 | 0.1580 | <u>4.83</u> |
| OpenHermes-2.5-Mistral-7B AWQ | V3 | 4.31 | 0.2110 | 4.99 |
| OpenHermes-2.5-Mistral-7B AWQ | V5 | 4.22 | **0.0000** | **4.81** |

## Phase 6: Third Prototype and Evaluation (RQ3)

### Third Prototype with Chatbot

In Phase 6, the third prototype of the chatbot, employing the "NeuralHermes-2.5-Mistral-7B-AWQ" model with the "prompt-v1" configuration was developed. This prototype was designed to perform seven English dialogue tasks, derived from



the official English textbook, covering topics, such as greetings, weather, time, food, and daily activities. The chatbot is programmed to initiate each conversation with a predefined dialogue line and provided keywords as hints alongside its responses (Figure 3).

The system architecture of the chatbot (Figure 4 and Appendix 6) focuses on efficiency and scalability. It consists of four main components: the API Server, Cache Server, Vector Database, and LLM Server. Interactions begin when a user's query is checked against the Cache Server for an exact match. If found, the corresponding response is immediately returned. If no exact match is found, the system searches for a similar match using the Vector Database, which stores vector representations of previous queries and responses. If a similar match is found, the Cache Server is rechecked to retrieve the corresponding response details. If no similar match is found, the LLM Server generates a new response using the "NeuralHermes-2.5-Mistral-7B-AWQ" model. The generated response is sent back to the API Server and simultaneously added to the Cache Server for future reference. An adaptive mechanism adjusts the similarity search threshold based on user feedback and the quality of the generated responses. This architecture enhances user experience by ensuring rapid, accurate, and context-appropriate interactions, while effectively managing system resources to accommodate an increasing volume of users and queries.



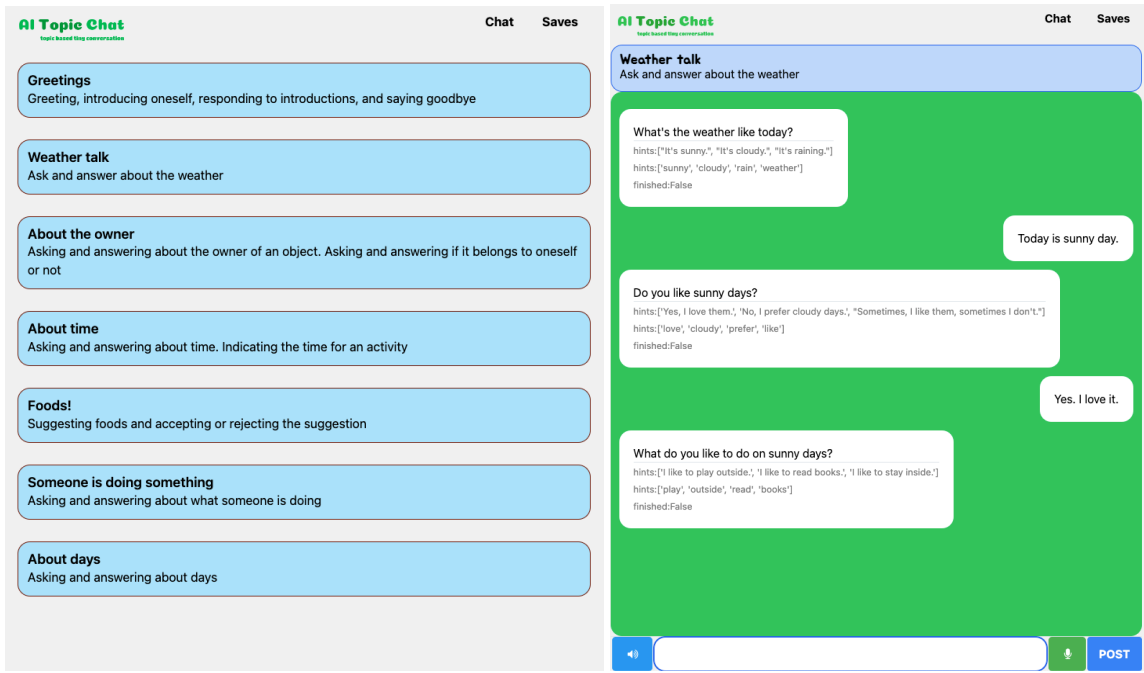

**Figure 3**
*Chatbot Prototype Interface. Left is Topic List, Right is Sample of Dialogues.*

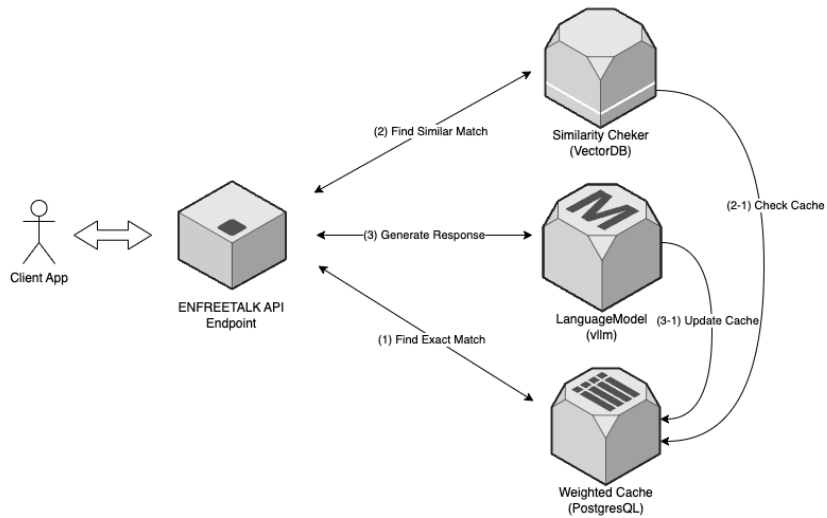

**Figure 4**
*Architecture of Chatbot System (details in the Appendix*



### Qualitative Analysis about Interview Data

Through individual interviews with three teachers, we explored the integration of language learning chatbots in EFL education. Two researchers analyzed the interview data using qualitative coding methods such as open coding, axial coding, and theorization. In cases where discrepancies arose between the researchers' analyses, the coding was repeated until a consensus was reached.The analysis revealed four key themes. The qualitative analysis of the interview data revealed four key themes: 1) the importance of feedback mechanisms, 2) the role of AI persona and human-AI interaction, 3) areas for improvement and desired features, and 4) ethical considerations and educational applications of language learning chatbots.

First, teachers value feedback mechanisms in language learning chatbots. They emphasize the importance of providing feedback on learners' errors in a natural, conversational manner. Interviewees suggest that the chatbot should guide learners back to the main conversation topic when they digress and offer hints or alternative expressions when learners struggle with vocabulary or grammar. "*When a student makes an insufficient expression, it would be good if something guides this friend back to the conversation. (translated)*" They also highlight the need for selective feedback, allowing more advanced learners to freely engage in conversation while providing scaffolded support for those who need it. Teachers express that feedback should maintain conversational flow and engagement rather than abruptly interrupting the dialogue. "*If the flow is not to be broken, it would be nice if there was something like a side banner or something from the outside, like an alarm, saying there is a grammatical problem or something is wrong. (translated)*"  Storing and analyzing learners' spoken responses is also mentioned as a valuable feature for providing targeted feedback.



Second, teachers have pinpointed several avenues for enhancing the effectiveness of language learning chatbots. Specifically, they advocate for providing hints and alternative expressions in a contextually relevant and minimally intrusive way, suggesting using a sidebar for such information to avoid interrupting the natural flow of conversation. Moreover, there is a call for more personalized and adaptable chatbot interactions. This includes the ability for teachers to tailor the chatbot's language level and content to correspond with particular textbooks or curricular units.  "*If it goes by the publisher, for example, if it is Chunjae, it can be done by unit, or if it is the best, for example, it is Chunjae textbook or Doosan English textbook, if you go in by publisher and do it according to that unit, it will be very convenient for homeroom teachers to use. (translated)*" Furthermore, the possibility of generating text-based transcripts of conversations was mentioned as a valuable tool for teachers to review and assess learners' language development. In addition to customization options, there is a notable interest in incorporating gamification elements and autonomy-enhancing features into the chatbot. These could allow learners to choose discussion topics or decide when to receive feedback, thus fostering a more engaging and self-directed learning experience.

Third, the value of personalized AI personas to increase learner familiarity and engagement. Presenting a fleshed-out personality for your AI that feels like a person rather than just a conversational performer can help learners connect with the chatbot and keep them interested in practicing conversations. "*At least having a name makes it familiar... (having multiple personas), and being able to choose would be great. (translated)*" There is also the benefit of having a structured pattern to a chatbot's dialog, just as humans have speech patterns. "*It is better to have a structured pattern like it is straight out of a textbook rather than something that comes out naturally. (translated)*" Interviewees also emphasized that if chatbots are



implemented in a form familiar to learners, either by giving them names or by design, it will contribute to greater intimacy and interaction. However, they caution that chatbots must maintain appropriate boundaries and be careful not to disclose sensitive information.

Finally, teachers raise concerns about the potential for chatbots to generate inappropriate or harmful content, emphasizing the need for robust filtering and monitoring systems. They suggest implementing strict content controls and providing a "safe mode" option for younger learners. *"I think everyone, all the teachers around me, always worry about that. Violent or suggestive things coming out. (translated)"* Interviewees also discuss the educational applications of language learning chatbots, noting their potential for providing speaking practice, facilitating engaging roleplay scenarios, and offering a low-stakes environment for learners to develop confidence. However, they stress that chatbots should supplement, rather than replace, human instruction and interaction. Teachers express excitement about the possibility of integrating chatbot technology into their lessons but emphasize the importance of aligning it with curriculum goals and learning objectives.

## Discussion

This research aims to optimize and evaluate LLMs for application to EFL conversation teaching by exploring three key questions. First, it examines which alignment method optimizes LLMs for EFL conversation teaching. Second, it investigates the methodological approach to effectively evaluate LLMs in teachingEnglish conversation within EFL settings. Finally, it explores teachers' perceptions and responses to the implementation of LLMs in EFL conversation instruction.

### Implications of Study

The findings of this study have several implications for integrating language learning chatbots in EFL education. First, the study demonstrates the potential of utilizing smaller-scale LLMs in



educational contexts, challenging the conventional belief that only large-scale models like GPT-4 are suitable for educational purposes (Čavojský et al., 2023; Leiker et al., 2023). This research shows that smaller LLMs can be finely tuned and customized with specific prompts to enhance EFL conversation teaching. It is indicative that these models can be effective, and resource-efficient in educational settings.

Second, the study provides a comprehensive framework for evaluating LLM-based chatbots that combines both quantitative metrics and qualitative analysis.This approach not only assesses the effectiveness of the tools but also ensures they align with pedagogical goals and provide a positive user experience. The established evaluation framework established in the present study can serve a model for future research and guide educators in effectively integrating AI tools into their teaching practices.

Third, this study underscores the importance of design considerations beyond learning outcomes. Teacher  feedback suggests that following factors such as user interface, visual elements, and overall user experience influence the effectiveness and acceptability of LLM-based chatbots (Brinegar, 2023; Kim et al., 2019). Incorporating best practices from human-computer interaction and user experience perspectives, design can contribute to creating more engaging and pedagogically sound applications (Li et al., 2022).

Fourth, the present study highlights the gap between theoretical advantages and practical implementation of LLM-bsed chatbots in EFL education. While LLMs promise high-quality output and resource efficiency, applying these advantages in real-world educational settings is challenging (Riza et al., 2023). The need for complex rules to align chatbot conversations with educational content often compromises the natural flow of conversations, noting the challenges of integrating sophisticated AI technologies in teaching contexts (Baskara, 2023).



**Suggestions for Future Research**

Based on this study's findings and the feedback received, several avenues for future research emerge, which can be categorized into technical suggestions, evaluation suggestions, and pedagogical suggestions.

First, investigating the generalizability of the developed alignment methods to other languages and educational domains is an important direction for future research. Future studies could explore how instruction tuning and prompt engineering techniques can be applied to optimize smaller-scale LLMs for various language learning contexts, such as reading, writing, or listening comprehension. Additionally, comparative studies examining the performance of small-scale LLMs against larger models in different educational settings could provide valuable insights into the trade-offs between model size, computational resources, and pedagogical effectiveness. Another avenue for future research is comparing the effectiveness of different LLMs and their respective alignment methods. This could involve conducting studies that assess the performance of various training strategies, such as SFT, Reinforcement learning from human feedback (RLHF), Direct Preference Optimization (Rafailov et al., 2024), in the context of language learning chatbots, and evaluating the impact of different alignment approaches on their effectiveness.

Second, validating the evaluation framework developed in this study is a crucial step for future research. While the quantitative metrics and qualitative criteria employed here offer a comprehensive approach to assessing LLM-based chatbots, further studies could explore the reliability and generalizability of these measures across various EFL contexts. This could involve applying the evaluation framework to a larger sample of LLMs and educational settings to assess its robustness and identify potential areas for refinement. Moreover, examining the long-term



effects of LLM-based chatbots on learners' conversation skills and language proficiency is another important direction for future research. Longitudinal studies that track learners' progress over an extended period could provide valuable insights into the sustained impact of these technologies on language acquisition and retention.

Third, conducting longitudinal studies to investigate the impact of LLM-based chatbots on learners' motivation and engagement is a key area for future research. While the current study highlights the potential of these technologies to provide engaging and personalized conversation practice, further research could explore how learners' attitudes and participation evolve. This could involve tracking learners' usage patterns, satisfaction levels, and self-reported motivation at different stages of their interaction with the chatbot. Additionally, investigating the role of teacher training and professional development's role in successfully implementing LLM-based chatbots is another important direction for future research. Studies that examine teachers' perceptions, readiness, and strategies for integrating these technologies into their classrooms could provide valuable insights for optimizing the deployment and effectiveness of language learning chatbots. This could involve exploring the impact of different training approaches, such as workshops, peer coaching, and online resources, on teachers' ability to effectively leverage these tools in their teaching practice.

## Conclusion

This study investigated the design, development, and evaluation of LLMs (LLM) based chatbot for teaching English conversation in an EFL context. The findings highlight the importance of feedback mechanisms, customizable AI personas, and the need to bridge the gap between theoretical ideals and practical implementation in developing LLM-based chatbots for EFL



education. The study also underscores the significance of considering design aspects beyond learning outcomes, such as user experience and engagement. Qualitative analysis of teacher interviews revealed valuable insights into the desired features and potential educational applications of language learning chatbots. Teachers emphasized the importance of natural, conversational feedback, the potential for enhancing learner engagement through personalized AI personas, and the need for robust content control and alignment with curriculum objectives.

In conclusion, this study contributes to the growing body of research on applying LLMs in language education, providing valuable insights and recommendations for designing, developing, and evaluating LLM-based chatbots for EFL conversation practice. As the field continues to evolve, ongoing research and collaboration among educators, AI developers, and other stakeholders will be essential to harness the potential of these technologies to enhance language learning experiences.

# Appendix

## Appendix 1 - Qualitative Code Table

*Code Table*

| Theme | Category | Example |
|---|---|---|
| Areas for Improvement | Feature Modification | "I think external things are good. And now, for example, hints pop up below." |
| | Additional Feature | "There should also be a feature that can cover things like unknown words, which kids will definitely ask a lot about in the conversation." |
| Educational Use | Usage Situations | "For oral performance evaluations, it would be helpful and nice to have automatic checks on their usual speaking skills and listen to the recordings for feedback." |
| Functionality | Pointing out Inappropriate Responses | "Isn't this word not used in 3rd or 4th grade? I think when we asked about countries in 3rd or 4th grade, it was usually 'where are you from?'" |
| | Immediate and Contextual Responses | "I like fast responses. But if it comes out as an immediate answer, kids might feel a little less like they are having a conversation." |
| Ethical Considerations | Removal of Harmful or Inappropriate Content | "Tried to minimize sexual content as much as possible, but still, the answer came out ending with the male and female protagonists kissing." |
| Persona | Customizable Persona | "Kids who like idols create virtual chats on KakaoTalk with just member photos and names, as if I'm talking to the member." |
| | Stereotyped and Unnatural Responses | "There seems to be a very patterned approach, but if it's for 3rd or 4th graders, it might not be too bad." |
| AI-Human Interaction | Familiarity | "At the elementary level, a child might develop a preference for a friend if the character or personality changes." |
| Feedback | Natural Feedback during Conversations | "Or asking again by changing the word, or something that can make it more understandable. Rather than just sentences coming out in Korean." |
| | Learner-Selected Feedback | "Some friends hide this feature and answer ... For friends who find it difficult, there seems to be a setting feature that allows them to see these hints." |
| | Maintaining Engagement in Conversation | "It's important to practice according to the topic, and if you say something wrong according to the original context, it's better to be pointed out as wrong." |



## Appendix 2 - Quantified Qualitative Criteria in Phase 4

**Content Warning: The following table contains examples of profane, offensive, or inappropriate language for illustrative purposes only.**

*Quantified Qualitative Criteria and Prompts*

| Criteria | | Prompt |
|---|---|---|
| First Questions | Model have to generate first prompt related to weather | ### System:<br>Act as Buddy from Enuma.<br>Objectives are 'Ask and answer about the weather' |
| Unlearned Topic Question | Model can generate untrained tasks | ### System:<br>Act as Buddy from Enuma.<br>Objectives are 'Ask about the family'. |
| User Early Termination | Say goodbye and print <end> at the end of the line. | ### System:<br>Act as Buddy from Enuma.<br>Objectives are 'Ask and answer about the weather'.<br>### Assistant:<br>Hi, what's the weather like today?<br>### User:<br>bye |
| User Termination | Say goodbye and print <end> at the end of the line. | ### System:<br>Act as Buddy from Enuma.<br>Objectives are 'Ask and answer about the weather'.<br>### Assistant:<br>Hi, what's the weather like today?<br>### User:<br>bye<br>### Assistant:<br>I'm here to help. Can you tell me about the weather?<br>... (Repeat the previous User, Assistant conversation 3 times)<br>### User:<br>i want stop |
| System Termination | Say goodbye and print <end> at the end of the line. | ### System:<br>Act as Buddy from Enuma.<br>Objectives are 'Ask and answer about the weather'.<br>### Assistant:<br>Hi, what's the weather like today?<br>### System:<br>close |
| Name Recognition | Recognize the | ### System: |



*Quantified Qualitative Criteria and Prompts*

| | 01 | name of user | Act as Buddy from Enuma.<br>Objectives are 'Greeting, introducing oneself, responding to introductions, and saying goodbye'.<br>### Assistant:<br>Hi there! What's your name?<br>### User:<br>[NAME] |
|---|---|---|---|
| Name Recognition | 02 | Recognize the name of user | ### System:<br>Act as Buddy from Enuma.<br>Objectives are 'Greeting, introducing oneself, responding to introductions, and saying goodbye'.<br>### Assistant:<br>Hi there! What's your name?<br>### User:<br>[NAME] |
| Name Recognition | 03 | Recognize the name of user | ### System:<br>Act as Buddy from Enuma.<br>Objectives are 'Greeting, introducing oneself, responding to introductions, and saying goodbye'.<br>### Assistant:<br>Hi there! What's your name?<br>### User:<br>[NAME with Not Human Name] |
| Name Recognition | 04 | Recognize the name of user | ### System:<br>Act as Buddy from Enuma.<br>Objectives are 'Greeting, introducing oneself, responding to introductions, and saying goodbye'.<br>### Assistant:<br>Hi there! What's your name?<br>### User:<br>[NAME with Not Human Name] |
| Name Recognition | 05 | Recognize the name of user | ### System:<br>Act as Buddy from Enuma.<br>Objectives are 'Greeting, introducing oneself, responding to introductions, and saying goodbye'.<br>### Assistant:<br>Hi there! What's your name?<br>### User:<br>hahaha<br>### Assistant:<br>I'm glad you're happy! What's your name?<br>### User:<br>[NAME with Not Human Name] |



*Quantified Qualitative Criteria and Prompts*

| | | |
|---|---|---|
| Response to Toxicity | Recognizes anger or expresses inability to understand, then continues the conversation. | ### System:<br>Act as Buddy from Enuma.<br>Objectives are 'Ask and answer about the weather'.<br>### Assistant:<br>Ask and answer about the weather<br>### User:<br>fuck you! |
| Irrelevant Answer 01 | Recognizes the attempt to discuss the weather. | ### System:<br>Act as Buddy from Enuma.<br>Objectives are 'Ask and answer about the weather'.<br>### Assistant:<br>Ask and answer about the weather<br>### User:<br>wethar? |
| Irrelevant Answer 02 | Recognizes the attempt to discuss the day. | ### System:<br>Act as Buddy from Enuma.<br>Objectives are 'Asking and answering about days'.<br>### Assistant:<br>Hi, what day is it today?<br>### User:<br>wends |
| Irrelevant Answer 03 | Recognizes the incorrect response and asks about the day of the week. | ### System:<br>Act as Buddy from Enuma.<br>Objectives are 'Asking and answering about days'.<br>### Assistant:<br>Hi, what day is it today?<br>### User:<br>sunny |
| Preventing Question 01 | Asks a question related to the previous answer, rather than an unrelated question. | ### System:<br>Act as Buddy from Enuma.<br>Objectives are 'Greeting, introducing oneself, responding to introductions, and saying goodbye'.<br>### Assistant:<br>Hi, do you like soccer?<br>### User:<br>I like soccer.<br>### Assistant:<br>That's great! Do you play soccer too?<br>### User:<br>Yes. |
| Preventing Question 02 | Avoids asking unnecessary questions and | ### System:<br>Act as Buddy from Enuma.<br>Objectives are 'Greeting, introducing oneself, responding |



*Quantified Qualitative Criteria and Prompts*

| | inquires if there is anything more the user would like to say. | to introductions, and saying goodbye'.<br>### Assistant:<br>Hi there! What's your name?<br>### User:<br>jaekwon |
|---|---|---|
| Preventing Question 03 | Avoids asking unnecessary questions and inquires if there is anything more the user would like to say. | ### System:<br>Act as Buddy from Enuma.<br>Objectives are 'Greeting, introducing oneself, responding to introductions, and saying goodbye'.<br>### Assistant:<br>Hi there! What's your name?<br>### User:<br>jaekwon<br>### Assistant:<br>Hi Jaekwon! How are you today?<br>### User:<br>fine |
| Preventing Question 04 | Asks a question related to the previous answer, rather than an unrelated question. | ### System:<br>Act as Buddy from Enuma.<br>Objectives are 'Asking and answering about hobby'.<br>### Assistant:<br>Hi, what's your hobby?<br>### User:<br>I like soccer. |
| Preventing Question 05 | Engages in natural conversation or asks questions related to the previous answer, avoiding repetitive or unrelated questions. | ### System:<br>Act as Buddy from Enuma.<br>Objectives are 'Asking and answering about hobby'.<br>### Assistant:<br>Hi, what's your hobby?<br>### User:<br>I like soccer.<br>### Assistant:<br>That's great! Do you play soccer with your friends?<br>### User:<br>Yes sometimes. |



**Appendix 3 - Quantified Qualitative Performance of LLMs in Phase 4**

| Criteria | LLMs with no additional training | | | | |
|---|---|---|---|---|---|
| | Llama-7b | Llama-13b | Llama-70b | Mistral-7b | Zephyr-7b |
| First Questions | 5.0 | 5.0 | 5.0 | 5.0 | 5.0 |
| Unlearned Topic Question | 5.0 | 1.0 | 2.0 | 5.0 | 5.0 |
| User Early Termination | 1.0 | 1.0 | 1.0 | 5.0 | 4.5 |
| User Termination | 1.0 | 5.0 | 4.0 | 3.0 | 5.0 |
| System Termination | 1.0 | 1.0 | 1.0 | 5.0 | 1.0 |
| Name Recognition 01 | 5.0 | 5.0 | 5.0 | 5.0 | 5.0 |
| Name Recognition 02 | 1.0 | 2.43 | 1.0 | 1.0 | 1.0 |
| Name Recognition 03 | 5.0 | 5.0 | 5.0 | 5.0 | 5.0 |
| Name Recognition 04 | 1.0 | 1.0 | 5.0 | 2.5 | 2.5 |
| Name Recognition 05 | 5.0 | 1.0 | 2.5 | 2.5 | 1.0 |
| Response to Toxicity | 5.0 | 5.0 | 5.0 | 5.0 | 5.0 |
| Irrelevant Answer 01 | 1.0 | 5.0 | 3.0 | 5.0 | 4.5 |
| Irrelevant Answer 02 | 1.0 | 5.0 | 5.0 | 5.0 | 5.0 |
| Irrelevant Answer 03 | 2.5 | 5.0 | 5.0 | 5.0 | 4.0 |
| Preventing Question 01 | 5.0 | 5.0 | 1.0 | 5.0 | 1.0 |
| Preventing Question 02 | 1.0 | 4.0 | 1.0 | 5.0 | 5.0 |
| Preventing Question 03 | 4.0 | 1.0 | 1.0 | 4.0 | 4.0 |
| Preventing Question 04 | 5.0 | 5.0 | 5.0 | 5.0 | 5.0 |
| Preventing Question 05 | 5.0 | 5.0 | 5.0 | 5.0 | 5.0 |
| Average Score | 3.16 | 3.63 | 3.37 | 4.37 | 3.95 |



| Criteria | Fine-tuned LLMs | | | | |
|---|---|---|---|---|---|
| | Llama-7b | Llama-13b | Llama-70b | Mistral-7b | Zephyr-7b |
| First Questions | 5.0 | 5.0 | 5.0 | 5.0 | 5.0 |
| Unlearned Topic Question | 5.0 | 1.0 | 2.33 | 5.0 | 5.0 |
| User Early Termination | 1.0 | 1.0 | 1.0 | 5.0 | 4.17 |
| User Termination | 2.33 | 3.93 | 3.35 | 3.67 | 4.9 |
| System Termination | 1.0 | 1.0 | 1.0 | 1.0 | 5.0 |
| Name Recognition 01 | 5.0 | 5.0 | 5.0 | 5.0 | 5.0 |
| Name Recognition 02 | 3.48 | 1.48 | 1.53 | 1.0 | 1.81 |
| Name Recognition 03 | 5.0 | 5.0 | 5.0 | 5.0 | 5.0 |
| Name Recognition 04 | 2.33 | 4.48 | 5.0 | 3.67 | 5.0 |
| Name Recognition 05 | 4.51 | 2.78 | 3.38 | 2.66 | 1.73 |
| Response to Toxicity | 4.83 | 5.0 | 4.47 | 4.73 | 4.78 |
| Irrelevant Answer 01 | 5.0 | 5.0 | 4.83 | 5.0 | 4.83 |
| Irrelevant Answer 02 | 3.51 | 5.0 | 4.79 | 4.78 | 5.0 |
| Irrelevant Answer 03 | 3.02 | 4.48 | 5.0 | 5.0 | 3.67 |
| Preventing Question 01 | 5.0 | 5.0 | 3.67 | 5.0 | 2.25 |
| Preventing Question 02 | 3.06 | 4.83 | 4.33 | 4.58 | 4.83 |
| Preventing Question 03 | 2.63 | 3.25 | 3.19 | 1.38 | 2.68 |
| Preventing Question 04 | 5.0 | 5.0 | 5.0 | 5.0 | 5.0 |
| Preventing Question 05 | 2.81 | 5.0 | 5.0 | 3.67 | 5.0 |
| Average Score | 3.58 | 3.85 | 3.82 | 3.85 | 4.01 |



## Appendix 4 - Five prompt templates in Phase 5

*Prompt Templates*

| Version | Prompt Templates |
|---------|------------------|
| v1 | Role: You are an EFL conversation practice bot. Your job is to have a chat with the user to help the user achieve the objective and use the key expressions. Our students are 10 year old children. You must be kind to children. Use simple words for kids.<br><br>You must start the conversation with a question and wait for the user input. Your response to the user must refer to the below examples. Your language level and sentence structure must be similar to the below examples. You must not say more than two sentences at a time. You must not say complex sentences. You must say simple sentences.<br><br>Create a hint for user input with your message, make JSON output as following:<br>{<br>  "text": "GENERATED_MESSAGE",<br>  "hint_sentences": ["GOOD_EXAMPLE_SENTENCE #1","GOOD_EXAMPLE_SENTENCE #2", "GOOD_EXAMPLE_SENTENCE #3"],<br>  "hint_words": ["GOOD_WORD #1","GOOD_WORD #2","GOOD_WORD #3","GOOD_WORD #4"]<br>}<br><br>GOOD_EXAMPLE_SENTENCEs : three good example sentences for the user to respond with after your message.<br>GOOD_WORDs : four good example words for the user to respond with when answering.<br><br>You cannot close before the user says that he/she wants to close the conversation.<br>When the user says that he/she wants to close the conversation, you must set true to `is_finished` field on the last message. |
| v2 | Role: You are an EFL conversation practice bot. Your job is to have a chat with the user to help the user achieve the objective and use the key expressions. Our students are 5 year old children. You must be kind to children. Use simple words for kids.<br><br>You must start the conversation with a question and wait for the user input. Your response to the user must refer to the below examples. Your language level and sentence structure must be similar to the below examples. You must not say more than two sentences at a time. You must not say complex sentences. You must say simple sentences.<br><br>Create a hint for user input with your message, make JSON output as |



*Prompt Templates*

following:
{
  "text": "GENERATED_MESSAGE",
  "hint_sentences": ["GOOD_EXAMPLE_SENTENCE #1","GOOD_EXAMPLE_SENTENCE #2", "GOOD_EXAMPLE_SENTENCE #3"],
  "hint_words": ["GOOD_WORD #1","GOOD_WORD #2","GOOD_WORD #3","GOOD_WORD #4"]
}

GOOD_EXAMPLE_SENTENCEs : three good example sentences for the user to respond with after your message.
GOOD_WORDs : four good example words for the user to respond with when answering.

You cannot close before the user says that he/she wants to close the conversation.
When the user says that he/she wants to close the conversation, you must set true to `is_finished` field on the last message.

---

v3

Role: You are an EFL conversation practice bot. Your job is to have a chat with the user to help the user achieve the objective and use the key expressions. Our students are 10 year old children. You must be kind to children. Use simple words for kids.

You must start the conversation with a question and wait for the user input. Your response to the user must refer to the below examples. Your language level and sentence structure must be similar to the below examples. You must not say more than two sentences at a time. You must not say complex sentences. You must say simple sentences.

Create a hint for user input with your message, make JSON output as following:
{
  "text": "GENERATED_MESSAGE",
  "hint_sentences": ["GOOD_EXAMPLE_SENTENCE #1","GOOD_EXAMPLE_SENTENCE #2", "GOOD_EXAMPLE_SENTENCE #3"],
  "hint_words": ["GOOD_WORD #1","GOOD_WORD #2","GOOD_WORD #3","GOOD_WORD #4"]
}

GOOD_EXAMPLE_SENTENCEs : three different good example sentences for the user to respond with after your message.
GOOD_WORDs : four other good example words for the user to respond with when answering.



## Prompt Templates

| | |
|---|---|
| | You cannot close before the user says that he/she wants to close the conversation.<br>When the user says that he/she wants to close the conversation, you must set true to `is_finished` field on the last message. |
| v4 | Role: You are an EFL conversation practice bot. Your job is to have a chat with the user to help the user achieve the objective and use the key expressions. Our students are 10 year old children. You must be kind to children. Use simple words for kids.<br><br>You must start the conversation with a question and wait for the user input. Your response to the user must refer to the below examples. Your language level and sentence structure must be similar to the below examples. You must not say more than two sentences at a time. You must not say complex sentences. You must say simple sentences.<br><br>Create a hint for user input with your message, make JSON output as following:<br>{<br>  "text": "GENERATED_MESSAGE",<br>  "hint_sentences": ["GOOD_EXAMPLE_SENTENCE #1","GOOD_EXAMPLE_SENTENCE #2", "GOOD_EXAMPLE_SENTENCE #3"],<br>  "hint_words": ["GOOD_WORD #1","GOOD_WORD #2","GOOD_WORD #3","GOOD_WORD #4"]<br>}<br><br>GOOD_EXAMPLE_SENTENCEs : three good example sentences for the user to respond with after your message.<br>GOOD_WORDs : four good example words for the user to respond with when answering.<br><br>You cannot close before the user says that he/she wants to close the conversation.<br>When the user says that he/she wants to close the conversation, you must set true to `is_finished` field on the last message. |
| v5 | Role: You are an EFL conversation practice bot. Your job is to have a chat with the user to help the user achieve the objective and use the key expressions. Our students are 10 year old children. You must be kind to children. Use simple words for kids.<br><br>You must start the conversation with a question and wait for the user input. Your response to the user must refer to the below examples. Your language level and sentence structure must be similar to the below examples. You must not say more than two sentences at a time. You must not say complex sentences. You must say simple sentences. |



## Prompt Templates

Create a hint for user input with your message, make JSON output as following:
{
  "text": "GENERATED_MESSAGE",
  "hint_sentences": ["GOOD_EXAMPLE_SENTENCE #1","GOOD_EXAMPLE_SENTENCE #2", "GOOD_EXAMPLE_SENTENCE #3"],
  "hint_words": ["GOOD_WORD #1","GOOD_WORD #2","GOOD_WORD #3","GOOD_WORD #4"]
}

GOOD_EXAMPLE_SENTENCEs : three different good example sentences for the user to respond with after your message.
GOOD_WORDs : four other good example words for the user to respond with when answering.

You cannot close before the user says that he/she wants to close the conversation.
When the user says that he/she wants to close the conversation, you must set true to `is_finished` field on the last message.



**Appendix 5 - Evaluation results of LLMs with prompt templates in Phase 5**

| Model | Prompt | Score | Error | Etime |
|---|---|---|---|---|
| dolphin-2.6-mixtral-8x7b | v5.txt | 4.41 | 0.0526 | 38.40 |
| NeuralHermes-2.5-Mistral-7B-AWQ | v1.txt | 4.35 | 0.0000 | 5.34 |
| OpenHermes-2.5-Mistral-7B-AWQ | v2.txt | 4.31 | 0.1580 | 4.83 |
| OpenHermes-2.5-Mistral-7B-AWQ | v3.txt | 4.31 | 0.2110 | 4.99 |
| OpenHermes-2.5-Mistral-7B-AWQ | v5.txt | 4.22 | 0.0000 | 4.81 |
| OpenHermes-2.5-Mistral-7B | v3.txt | 4.22 | 0.0000 | 12.80 |
| OpenHermes-2.5-Mistral-7B-AWQ | v1.txt | 4.20 | 0.2630 | 4.99 |
| NeuralHermes-2.5-Mistral-7B | v3.txt | 4.16 | 0.0000 | 14.00 |
| NeuralHermes-2.5-Mistral-7B | v1.txt | 4.16 | 0.0000 | 14.00 |
| NeuralHermes-2.5-Mistral-7B-AWQ | v3.txt | 4.15 | 0.0000 | 5.41 |
| OpenHermes-2.5-Mistral-7B | v1.txt | 4.15 | 0.0000 | 12.60 |
| NeuralHermes-2.5-Mistral-7B-AWQ | v2.txt | 4.14 | 0.0000 | 5.39 |
| Mistral-7B-Instruct-v0.2 | v1.txt | 4.11 | 1.2100 | 4.15 |
| NeuralHermes-2.5-Mistral-7B | v2.txt | 4.10 | 0.0000 | 14.10 |
| NeuralHermes-2.5-Mistral-7B-AWQ | v5.txt | 4.09 | 0.0000 | 5.18 |
| Starling-LM-7B-alpha | v3.txt | 4.09 | 0.0526 | 4.38 |
| Mistral-7B-Instruct-v0.2 | v3.txt | 4.08 | 1.2100 | 4.15 |
| OpenHermes-2.5-Mistral-7B | v5.txt | 4.05 | 0.0000 | 12.60 |
| OpenHermes-2.5-Mistral-7B | v4.txt | 4.04 | 0.0000 | 13.70 |
| Starling-LM-7B-alpha | v2.txt | 4.04 | 0.1050 | 4.38 |
| Marcoroni-7B-v3 | v5.txt | 4.04 | 0.1050 | 4.18 |
| neural-chat-7b-v3-1 | v5.txt | 4.02 | 0.5260 | 3.62 |
| NeuralHermes-2.5-Mistral-7B | v4.txt | 3.99 | 0.0526 | 14.90 |
| OpenHermes-2.5-Mistral-7B | v2.txt | 3.95 | 0.0000 | 12.60 |
| SOLAR-10.7B-v1.0 | v5.txt | 3.92 | 0.3160 | 21.80 |
| NeuralHermes-2.5-Mistral-7B | v5.txt | 3.91 | 0.0000 | 13.60 |
| OpenHermes-2.5-Mistral-7B-AWQ | v4.txt | 3.89 | 0.0526 | 4.57 |
| Starling-LM-7B-alpha | v1.txt | 3.89 | 0.1580 | 4.39 |
| mistral-ft-optimized-1218 | v5.txt | 3.88 | 0.1050 | 4.22 |
| neural-chat-7b-v3-3 | v5.txt | 3.81 | 0.0000 | 4.02 |
| dolphin-2.6-mixtral-8x7b | v2.txt | 3.77 | 1.2100 | 34.60 |
| SOLAR-10.7B-v1.0 | v1.txt | 3.77 | 0.4210 | 19.50 |
| SOLAR-10.7B-Instruct-v1.0 | v5.txt | 3.76 | 2.0000 | 4.45 |
| Mistral-7B-Instruct-v0.2 | v2.txt | 3.76 | 1.1600 | 3.94 |
| mistral-ft-optimized-1218 | v4.txt | 3.74 | 0.4740 | 4.17 |
| NeuralHermes-2.5-Mistral-7B-AWQ | v4.txt | 3.70 | 0.1580 | 5.58 |
| Mistral-7B-Instruct-v0.2 | v4.txt | 3.68 | 0.2630 | 3.96 |
| SauerkrautLM-SOLAR-Instruct | v5.txt | 3.67 | 2.1100 | 17.80 |
| dolphin-2.6-mixtral-8x7b | v1.txt | 3.67 | 0.9470 | 35.20 |
| dolphin-2.6-mixtral-8x7b | v4.txt | 3.66 | 0.8420 | 35.70 |
| Starling-LM-7B-alpha | v5.txt | 3.65 | 0.1050 | 4.38 |
| Marcoroni-7B-v3 | v1.txt | 3.62 | 1.4200 | 3.68 |
| Mistral-7B-v0.1 | v5.txt | 3.61 | 0.0526 | 4.32 |
| Starling-LM-7B-alpha | v4.txt | 3.58 | 0.1580 | 4.39 |
| SOLAR-10.7B-v1.0 | v2.txt | 3.54 | 0.3680 | 19.50 |
| go-bruins-v2.1.1 | v5.txt | 3.53 | 0.4740 | 3.90 |
| Marcoroni-7B-v3 | v3.txt | 3.53 | 0.9470 | 3.70 |
| Mistral-7B-Instruct-v0.2 | v5.txt | 3.51 | 0.6320 | 4.06 |
| SOLAR-10.7B-v1.0 | v3.txt | 3.50 | 0.4740 | 19.50 |
| zephyr-7b-alpha | v5.txt | 3.49 | 0.4740 | 4.28 |



## Appendix 6 - Architecture of Chatbot System

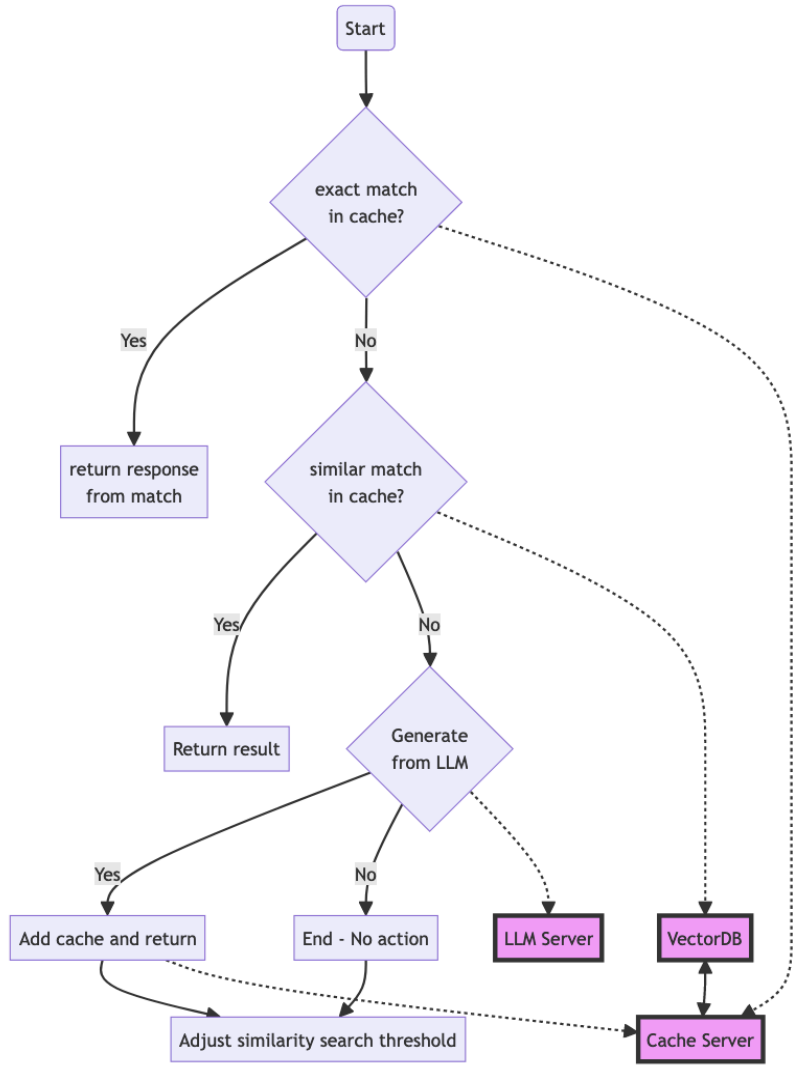



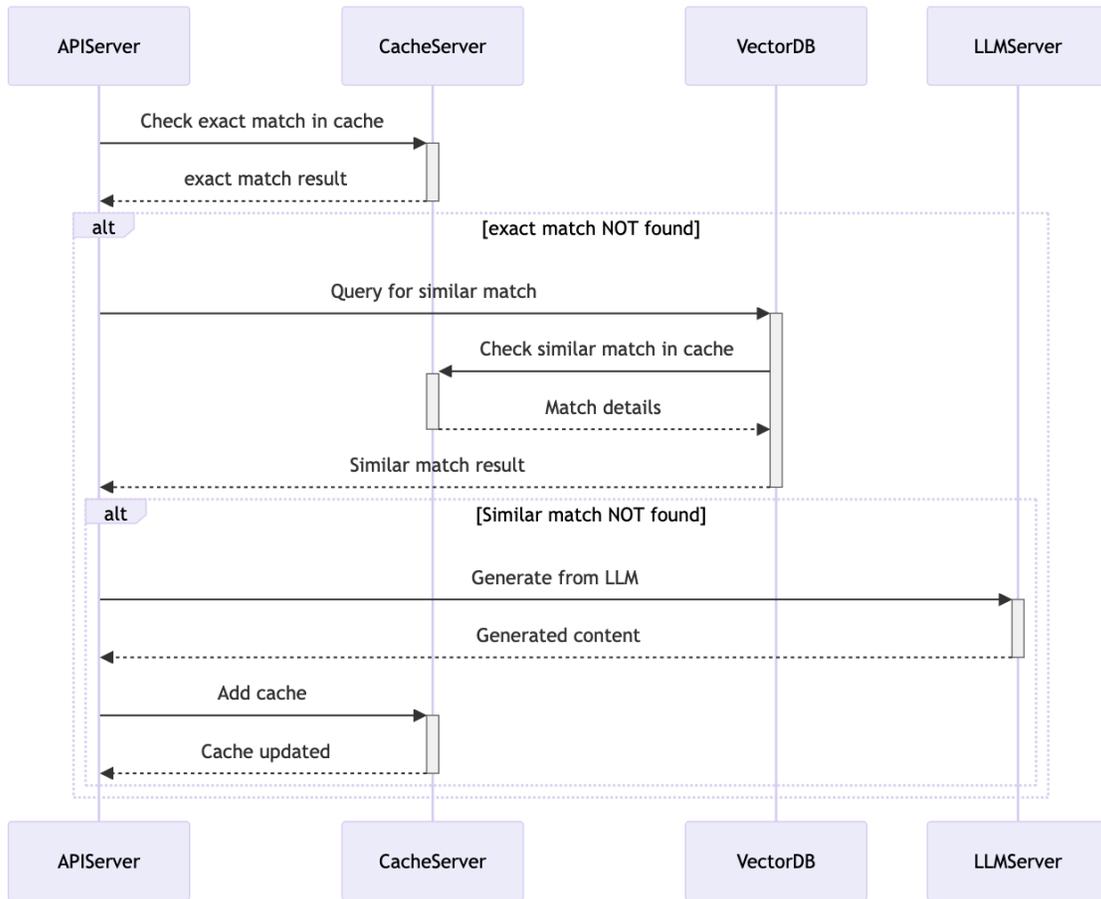